# Performance of TOF-RPC for the BGOegg experiment

**N. Tomida**[a,*]**, N. Tran**[a]**, M. Niiyama**[b]**, H. Ohnishi**[c,a] **and N. Muramatsu**[d]

[a] *Research Center for Nuclear Physics (RCNP), Osaka University,*
*Ibaraki, Osaka 567-0047, Japan*

[b] *Department of Physics, Kyoto University,*
*Kyoto 606-8502, Japan*

[c] *RIKEN (The Institute of Physical and Chemical Research),*
*Wako, Saitama 351-0198, Japan*

[d] *Research Center for Electron Photon Science (ELPH), Tohoku University,*
*Sendai, Miyagi 982-0826, Japan*

*E-mail:* natsuki@rcnp.osaka-u.ac.jp

ABSTRACT: We constructed a new time-of-flight (TOF) detector consisting of resistive plate chambers (RPCs) to measure particle energy in the BGOegg experiment. The BGOegg-RPC has a unique feature which enables us to cover a large area with a small number of readout channels. For this purpose, we developed the RPC with a strip size of 2.5 cm × 100 cm. The BGOegg-RPC covers an area of 320 cm × 200 cm with only 256 channels of readout electronics. In case of large readout RPCs, an originated signal is distorted and dispersed during propagation. In addition, there happens a signal reflection at the end of the strip. Although we designed the BGOegg-RPC and front-end electronics with minimized signal reflection, a small reflection still remained, deteriorating the resulting time resolution. After establishing calibration and correction methods to improve the performance of the BGOegg-RPC, we obtained the time resolution of $\sigma \sim 60$ ps around the central region of strips.

KEYWORDS: Resistive-plate chambers; Instrumentation and methods for time-of-flight (TOF) spectroscopy; Timing detector.

---

[*] Corresponding author.

# Contents

## 1. RPC with large readout strip

### 1.1 Introduction

A resistive plate chamber (RPC) is a gas counter made of parallel resistive plates [1]. Avalanches generated in the gaps induce pulse signals at the readout strip. The timing fluctuation of the avalanche development in the narrow-gap RPC is very small; σ~25 ps for 260 μm × 10 gap RPCs and σ<10 ps for 140 μm × 24 gap RPCs [2,3]. However, the time resolution of most of narrow-gap RPCs are worse than the time fluctuation of avalanches. The other sources which contribute to the overall time resolution of the TOF system are the timing jitter of front-end electronics, the TDC resolution, the hit position ambiguity in the readout strip, and the signal distortion during the propagation in the readout strip. We wished to construct the RPCs with a large single readout-strip, resulting cheap costs for the TDCs. However, the signal distortion during propagation deteriorates the time resolution in case of large readout strips.

## 1.2 Design of high time-resolution RPCs with large readout strips

There are four important problems to be solved in designing a high time-resolution RPC with large readout strips: 1) minimizing the effect of hit position ambiguities using the mean timing of signals at both ends of the readout strip, 2) minimizing signal distortions using the strip-type readout instead of the pad-type readout, 3) avoiding signal reflections at the end of strip by matching impedance between the readout strip and the amplifier, and 4) minimizing modal dispersions of signals by balancing the capacitive and inductive coupling of the RPC (electrostatic compensation). We describe how to deal with these problems in the following.

### 1.2.1 Minimizing the hit position ambiguity

Although single-end readout pads have been widely used for early TOF-RPCs in the experiments such as HARP, ALICE and STAR [4-6], the timing deterioration due to hit position ambiguities was problematics. The signal propagation velocity in the readout strip is about 50 ps/cm. Thus, small variation of the hit position can make the time resolution worse. The ALICE-RPC (pad: 2.4 cm $\times$ 3.7 cm) has an excellent time resolution of 50 ps when a 1 cm $\times$ 1 cm trigger is used [5]. However, it becomes worse to 85 ps when particles hit the whole pad [7]. Thus, both procedures to read signals at both ends of the readout strip, and to use the average timing of them to cancel the position ambiguity effect are important to achieve a better time resolution. For example, the BES-III RPC (pad: 2.4 cm $\times$ 9.1-14.1 cm) achieves a time resolution around 60 ps using the mean timing of signals at both ends under the condition where particles hit the whole pads [8].

### 1.2.2 Minimizing signal distortion during propagation

The time resolution of the pad-type RPCs becomes worse with increasing the pad area due to both the capacitance increase and the distortion of signal shapes resulting from different propagation paths [9]. The strip-type readout whose width is small enough compared to the strip length has less distortion of signals during their propagations. We tested a 2.5 cm x 108 cm strip and a 5.0 cm x 108 cm strip, and concluded that 2.5 cm-wide strip provides us a much better resolution for the BGOegg-RPC [10].

### 1.2.3 Avoiding the signal reflection

When the impedances of the readout strip and the amplifier are not matched well, a signal is partially reflected at the contact point of the strip and the amplifier. When the reflection signal overlaps with the prompt signal, the prompt signal is distorted and the time resolution becomes worse. Although the BGOegg-RPC is designed to minimize the reflection, a small reflection has still remained [10]. Problems associated with such reflections are observed in case of the BGOegg-RPC and are explained in the next sections.

### 1.2.4 Minimizing modal dispersion

The signal propagation in the readout strip can be described by the transmission line theory [11]. According to the theory, there are mainly two different propagation modes. In most cases, two modes have different propagation velocities causing dispersion of the signal during long propagation (modal dispersion). In such a case, the transmission line calculation shows slow rising time of the signal after long propagation. In addition to the signal shaping, the cross-talk level becomes large if the RPC is not designed properly. By adding extra layers of insulating materials to balance the capacitive and inductive coupling, the two velocities can be almost the same and the cross-talk can be reduced (electrostatic compensation). The BGOegg-RPC is not well designed to achieve this electrostatic compensation. Thus, we suffer a large cross-talk. On the other hand, the direct comparison of the signal rising time with the time resolution is difficult, since the time resolution is affected by many factors. These issues about the transmission and the compensation are discussed in more details in ref. [5].

**1.3 The BGOegg experiment**

The BGOegg experiment has been conducted at the LEPS2 beamline at SPring-8 in Japan since 2013. At the LEPS2 beamline, a photon beam up to 3.0 GeV is generated by backward Compton scattering of UV lasers off 8-GeV electrons in the SPring-8 storage ring. The photon energy resolution of the BGOegg calorimeter, consisting of 1320 BGO crystals, is 1.3% for 1 GeV gamma rays. The world's best resolution is realized for meson decay studies. The TOF-RPC is located at 12.5 m downstream of the target. The TOF-RPC is used for the energy measurement of particles at very forward direction. The TOF-RPC has been stably functioning in the BGOegg experiment since 2014.

**1.4 TOF-RPC**

The TOF-RPC covers an area of 320 cm (W) × 200 cm (H). It consists of 32 RPC modules of 25 cm (W) × 100 cm (H). The cross section of the module is shown in figure 1. It is a double stack RPC with 5 gaps of 260 μm width. The applied voltage across five gaps is 13.5 kV. A schematic drawing of the strip-end is shown in figure 2. There are eight strips in a module. A readout strip is 2.55 cm (W) × 100 cm (H). The strips are printed on PCB boards. The gap between strips is 0.5 mm. The signal is read at both top- and bottom-ends. A mixture gas of 90% $C_2H_2F_4$ (R134a), 5% $SF_6$ and 5% $C_4H_{10}$ (butane) is used at a flow rate of 300 cc/min.

Pre-amplifiers are connected to the readout strips using short copper lines (see figure 3). The anode strip is connected to the amplifier input and the cathode strip is connected to ground. The impedance of the anode strip is estimated to be 30~40 Ω from calculation using APLAC simulation and measurement using pulse generator. The input impedance of the amplifier is set to be 40 Ω. A small reflection remained even after this impedance matching [10]. This may be owing to the non-uniform shape of the strip-end or mismeasurement of the strip impedance.

We use "SUM" and "OR" of signals from two strips to reduce the number of ADC/TDC channels. For ADC, the signals from two neighboring strips are summed up at the amplifier and read by one channel of the ADC. For TDC, the signals from two strips in different chambers are "OR"ed at the "OR board" after the discrimination and read by one channel of the TDC. Thus, we need an algorithm to select the correct hit strip. More details of the detector and front-end electronics are described in ref. [12].

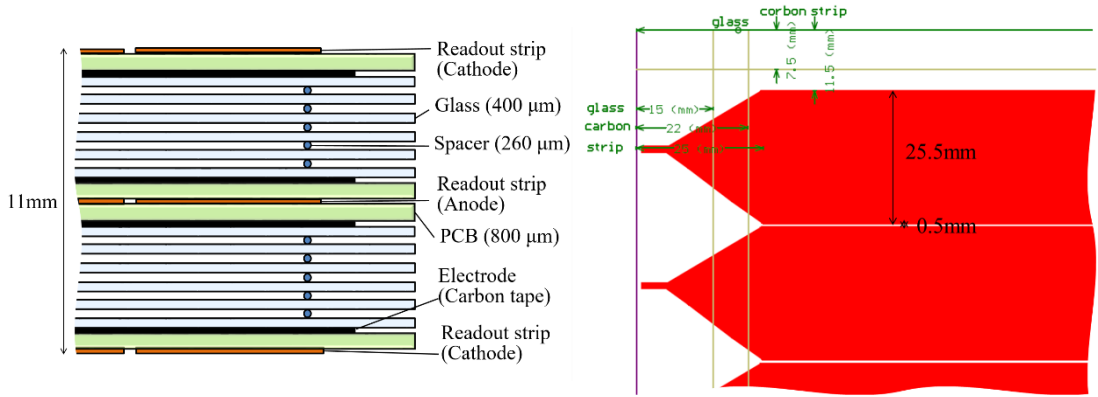

**Figure 1**. The cross section of the BGOegg-RPC.

**Figure 2**. The shematic drawing of the end of the strip.

The RPC time resolution mostly contributes to the total TOF time resolution. The time resolution of the TDC is about 30 ps. The start timing of the TOF is determined by the RF signal of the SPring-8 storage ring. The timing jitter of the RF signal is about 4 ps. The electron-beam bunch-width in the storage ring depends on the bunch filling pattern and is less than 14 ps. Thus, the start timing is determined with the resolution about 15 ps.

## 2. Signal shapes and comparison with the transmission line calculation

The waveforms of signals of the BGOegg-RPC triggered by cosmic rays are measured at different positions using an oscilloscope. The schematic drawing of the set-up is shown in figure 4. The hit position was determined as 1 cm × 1 cm by two finger scintillators placed on strip 4. The waveforms are shown in figure 5(a)-(c). The orange lines and the blue lines are the signal of strip 4 at the bottom-end (y=0 cm) and the top-end (y=100 cm), respectively. The cross-talk pulses are induced on the neighboring strip (strip 3) and shown by the pink (bottom) and the green (top) lines. The signal amplitude is normalized to the bottom-end signal as unit amplitude. The pulse height of the cross-talk is 20-40% of the signal.

Figure 5(a) shows the waveforms at y=50 cm. The waveforms at the top-end and the bottom-end are symmetric. The reflections are observed about 5 ns later from the prompt signals. The reflection observed at the top-end was generated at the bottom-end and travelled 100 cm at 50 ps/cm and arrived at the top-end. Figure 5(b) shows the waveforms at y=25 cm. The interval between the prompt signal and the reflection is about 2.5 ns for the top-end (the blue line) and 7.5 ns for the bottom-end (the orange line). Figure 5(c) shows the waveforms at 1 cm. In the case of the top-end (the blue line), the timing difference of the prompt signal and the reflection is about 0.1 ns. Thus, they are overlapped and cannot be distinguished. The prompt signal is distorted by the reflection, leading a worse time resolution. In addition, the pulse height becomes larger since the signal appears to be the sum of the prompt signal and the reflection. It leads faster signal timing compared to signals at around the center of the strip with the same total charge owing to slewing effect. This indicates necessity of different slewing correction at the different position along strip.

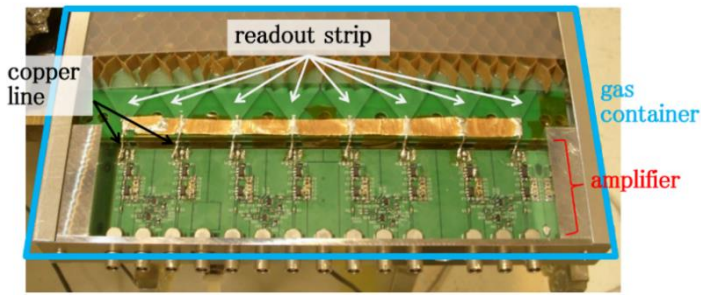

**Figure 3**. The amplifier and the connection point to the readout strip.

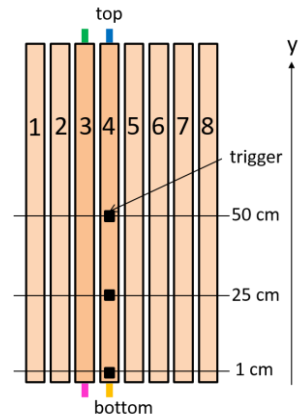

**Figure 4**. The set-up of the waveform measurement.

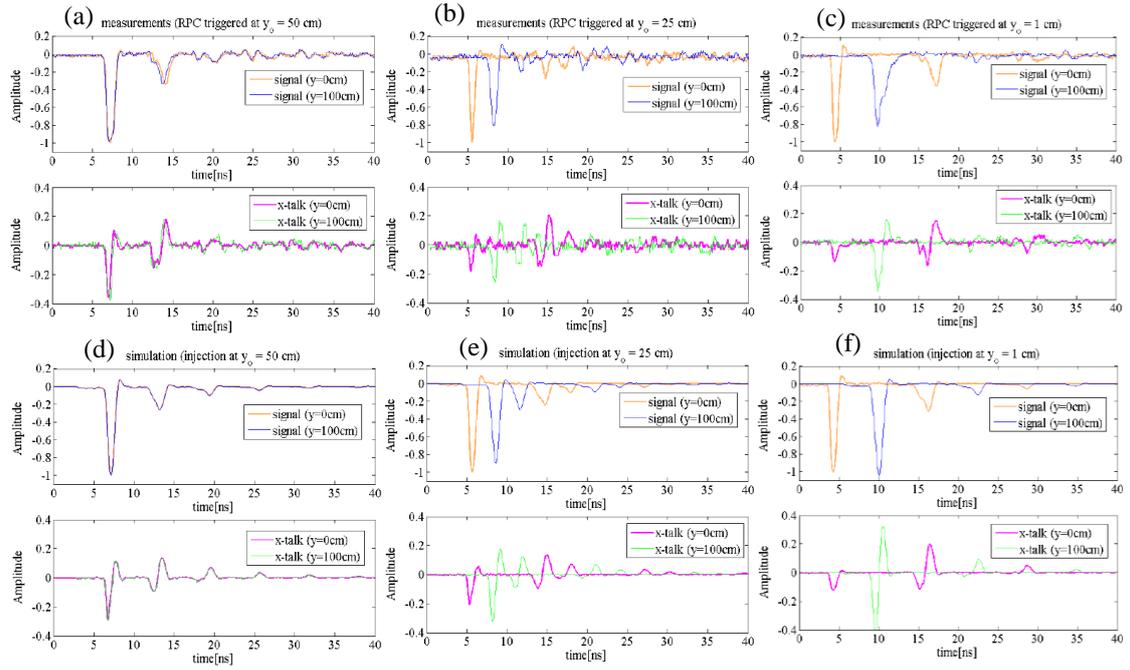

**Figure 5**. (a)-(c) : The measured signal at the trigger position of y=50 cm (a), 25 cm (b) and 1 cm (c). (d)-(f) : The simulation signal calculated based on the work of ref. [5] at the signal generation point of y=50 cm (d), 25 cm (e) and 1 cm (f). The orange lines are the signal observed at the bottom-end (y=0 cm). The blue lines are at the top-end (y=100cm). The pink and green lines are the siganls observed at the neighboring strip (cross-talk).

The results from the transmission line calculation are shown in figure 5(d)-(f). The geometry of the BGOegg-RPC is used as the input parameter. The triangle-shape at the end of the strip is not reproduced in the calculation and it is assumed to be rectangle. The calculation well describes what is observed in the cosmic-ray test including cross-talk. This demonstrates the validity of the transmission line calculation. Small disagreement is observed in the signal at the bottom-end (the blue line) at y=1 cm. The prompt signal and the reflection are fully overlapped in the simulation, although not in the data. This may be caused from the difference of the strip-end shape between the detector and the simulation.

## 3. RPC calibration

For the analysis of the BGOegg-RPC, the following calibrations and corrections are performed,
1) hit strip selection,
2) y-position calibration,
3) slewing correction,
4) t0 correction.

A lot of electron-positron showers come to the TOF-RPC. Since they travel at constant light speed, we used the shower events for the calibration and the evaluation of the time resolution. The details are explained in the following.

## 3.1 Hit strip selection

As explained in section 1.4, the signals from two strips are summed/ORed, and are read by one ADC/TDC channel. In addition, the BGOegg-RPC suffers large cross-talk. Thus, we have developed an algorithm to find the correct hit strip. First, the strips with the largest charge in the chamber are searched for. At this moment, two strips using a common ADC channel are chosen. Second, the hit timings of the two strips are compared. Though both strips have TDC hits, the pulse at the hit strip is larger than those observed at the neighboring strip (the cross-talk strip) as can be seen in figure 5. The arrival timing of the pulse at the cross-talk strip is found to delay by about 200 ps due to time-walk effect. Thus, the strip which has a faster timing is selected as the hit strip.

## 3.2 Y-position calibration

The hit position along strip (y-position) is derived from the difference of the timing at the top-end and the bottom-end. However, the time difference is not linearly scaled as a function of the y-position. The signal timing is faster near the end of the strip since the pulse height is larger because of the overlap of the prompt signal and the reflection as seen in figure 5(c).

In order to calibrate the y-position including this effect, we took a special run using finger scintillators. Finger scintillators whose width of 1 and 2 cm are set at y=0 cm (bottom-end), 20 cm, 40 cm, 60 cm, 80 cm and 100 cm (top-end) of the strip. Figure 6 shows the time difference of the top-end and the bottom-end when one of scintillators has hit. There are more particles in right side since the bottom side is close to the beam height (0 cm). The vertical dot lines are drawn every 2.55 ns which is the interval of the middle two peaks. The intervals of the peaks are shorter near the end of the strip compared to the middle of the strip. The difference from the standard lines are fitted using a 3rd-order polynomial function. The obtained function is used to calculate correct y-position.

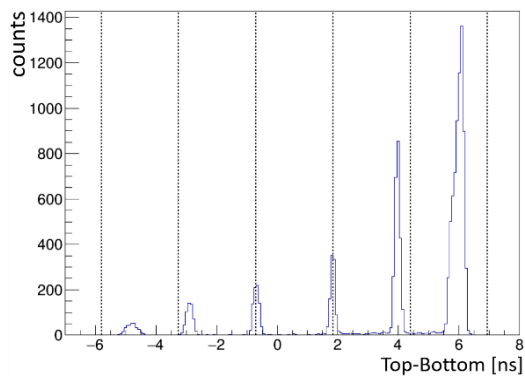

**Figure 6**. The time difference of the top-end and the bottom-end when one of the finger scintillators has hit. The vetrical lines are drawn every 2.55 ns.

## 3.3 Slewing correction

Figure 7(a) show the scatter plot of the average of the timing at the top-end and the bottom-end before slewing correction versus the y-position. First, the timing is found to delay at the large y-position. This is because of the flight length difference at different y-positions. The flight length from the target is about 4 cm longer at y=100 cm compared to at y=0 cm. This corresponds to the arrival time difference about 140 ps when particles flight at the speed of light. Second, timing is fast near the end of the strip. This is because of the signal reflection. As discussed in section 2 and 3.2, reflections overlap with the prompt signal near the end of strip. Therefore, the pulse height becomes large, and it causes a faster signal arrival due to the time-walk effect.

Because of the reflection and dispersion effects, the time-walk is different at various positions even within a single strip. Thus, we classified the signals from a single strip into those from each subdivided 5 cm y-region and made slewing corrections for signals from each region. The used fitting function was $C_0 + C_1/\sqrt{ADC} + C_2/ADC$, where ADC is the sum of the charge of the top-end and the bottom-end, and $C_0$, $C_1$ and $C_2$ are the fitting parameters. Figure 7(b) shows the timing distribution along y-position after the slewing corrections. Both the flight length effect and the reflection effect are corrected.

### 3.4 T0 correction

Figure 8 shows the peak position of the timing (RPC-RF) of electron events after slewing corrections along with each run number. The data taking was carried out for about 200 days per year. We picked up data of ten days as a sample. To clearly indicate the day-night effects, the vertical lines in the figure indicate mid-night data points. The peak position shifted about 30 ps together with temperature change. This is because the cable length of the RF signal slightly changes depending on the temperature. The RF cable is guided from the storage ring building to the LEPS2 building. Thus, the temperature cannot be controlled outside of the buildings. There was rain around run8000-8050 and the temperature change was small. The peak position is obtained every run and its fluctuation is corrected.

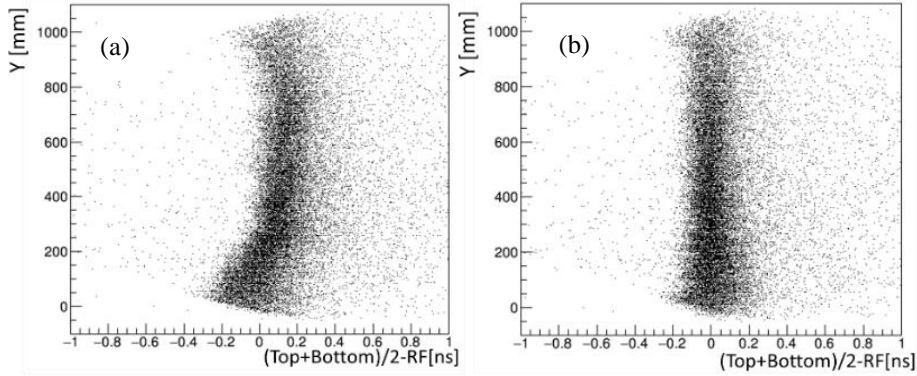

**Figure 7**. The scatter plot of the average of the timing at the top-end and the bottom-end versus the y-position before slewing correction (a) and after the correction (b).

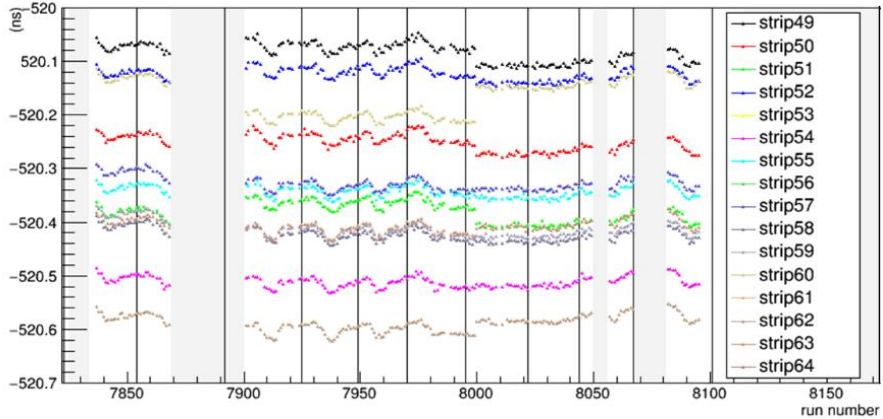

**Figure 8**. The run dependence of the mean position of the timing peak of electron events. The verical lines inserted are drawn at mid-night data points. The regions shadowed in the gray color are junk runs. About 30 ps day-night shift is observed due to the timing shift of the RF signal. Run8000-8050 were taken in rainy days.

## 4. RPC performance

### 4.1 Special data using finger scintillators

To examine the position dependence of the time resolution of the top-end and the bottom-end separately, we took the hit position data. By defining the hit position using finger scintillators, we could evaluate the performance of each end, free from the position ambiguity effect. The width of finger scintillators was 1 cm and 2 cm. They are put in front of a module at y=0 cm, 20 cm, 40 cm, 60 cm, 80 cm and 100 cm. Figure 9 shows the time resolution of the corresponding strip at each position. The time resolution is obtained by a Gaussian fit of the time distribution of RPC-RF. The red line is the time resolution of the mean of the top- and bottom-ends. The blue line is the time resolution for the top-end, and the green is for the bottom-end. The number of events at y=100 cm is small and the slewing correction does not work well. The time resolution is worse at the far-end where the prompt signal and the reflection are overlapped. The mean time resolution in the central region of the strip is σ=60~65 ps and slightly worse than that of σ=50 ps in the test experiment of a single module before the installation [12].

### 4.2 Performance during physics run

Figure 10 shows the y-position dependence of the time resolution for two chambers during physics data taking. Electron showers are used to obtain the time resolution. The y-position is obtained by the method described in section 3.2 and sliced every 10 cm. The resolution around the center of the strip is about σ = 60 ps. The resolutions are observed to be poor near the end of strips.

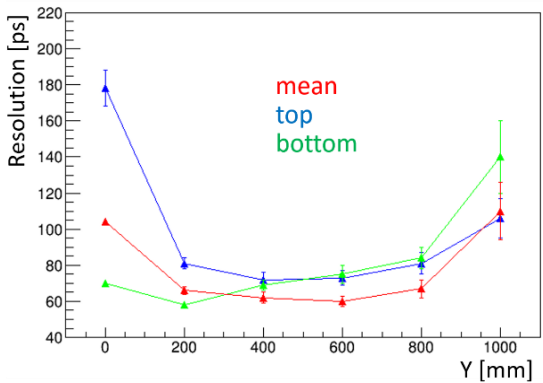
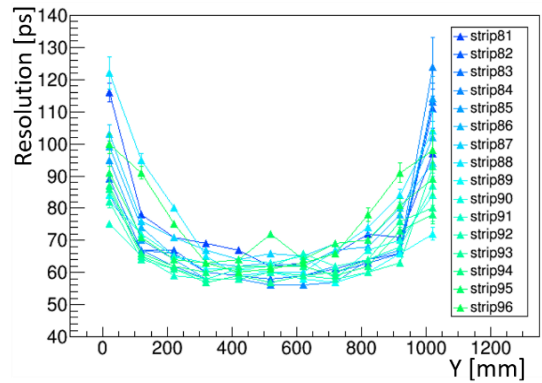

**Figure 9**. The time resolution of the mean (red), the top-end (blue) and the bottom-end (green) at y=0 cm, 20 cm, 40 cm, 60 cm, 80 cm and 100 cm. The y-position is defined using finger scintillators.

**Figure 10**. The time resolution of sixteen strips in two chambers. It is about 60 ps in the central region and is worse near the end of strips.

## 5. Summary

Large strip RPCs are desired for a reasonable TOF detector. We summarized important issues for the design of a large readout strip RPC. We showed the effectiveness of the transmission line calculation for describing the signal propagation in the readout strip of the RPC.

The BGOegg-RPC is designed to minimize some characteristic problems in use of large strips. However, the impedance matching of the BGOegg-RPC is not perfect and a small reflection still remains. We found that the reflection deteriorates the RPC performance. The established correction methods for the reflection effect is explained. The time resolution around 60 ps is obtained in the central region of the strips. The time resolutions near the end of strips are worse than the central region due to the reflection.

## Acknowledgments

We thank Dr. D. Gonzalez-Diaz for the advice and the calculation based on the transmission line theory. This research was supported by Japan MEXT/JSPS KAKENHI Grant number 24105711, 24608 and 251348, and the Ministry of science and Technology of Taiwan Grant number 103-2112-M-001-026-MY3.